\pdfoutput=1
\documentclass[12pt,a4paper]{article}
\usepackage[utf8]{inputenc}
\usepackage{a4wide}
\usepackage{amsmath,amssymb,amsfonts}
\usepackage{subfig,graphicx}
\usepackage[square,numbers,sort&compress]{natbib}
\usepackage[dvipsnames]{xcolor}
\usepackage[colorlinks,allcolors=blue]{hyperref}
\usepackage{dsfont}

\newcommand{\beq}{\begin{eqnarray}}
\newcommand{\eeq}{\end{eqnarray}}
\newcommand{\non}{\nonumber\\}
\DeclareMathOperator{\U}{U}
\DeclareMathOperator{\SU}{SU}

\newcommand{\p}{\partial}
\renewcommand{\i}{\mathrm{i}}
\renewcommand{\d}{\mathop{}\!\mathrm{d}}

\DeclareMathOperator{\Tr}{Tr}

\DeclareMathOperator{\MeV}{MeV}
\DeclareMathOperator{\fm}{fm}

\newcommand{\calA}{\mathcal{A}}
\newcommand{\calE}{\mathcal{E}}
\newcommand{\calF}{\mathcal{F}}

\newcommand{\wha}{\widehat{a}}
\newcommand{\bchi}{\boldsymbol{\chi}}
\newcommand{\btau}{\boldsymbol{\tau}}

\begin{document}
\pagenumbering{Roman}
\begin{titlepage}
  \begin{flushright}
    October 2023
  \end{flushright}
  \vskip3cm
  \begin{center}
    {\Large\bf Neutron stars in the Witten-Sakai-Sugimoto model}\\[2cm]
    {\bf Lorenzo Bartolini, Sven Bjarke Gudnason}\\[2cm]
    {Institute of Contemporary Mathematics, School of Mathematics and Statistics,\\ Henan University, Kaifeng, Henan 475004, P. R. China}
  \end{center}
  \vfill
  \begin{abstract}
    We utilize the top-down holographic QCD model, the
    Witten-Sakai-Sugimoto model, in a hybrid setting with the SLy4,
    soft chiral EFT and stiff chiral EFT
    equations of state to describe neutron stars with high precision.
    In particular, we employ a calibration that bootstraps the nuclear
    matter by fitting the Kaluza-Klein scale and the 't Hooft coupling
    such that the physical saturation density and physical symmetry
    energy are achieved.
    We obtain static stable neutron star mass-radius data via the
    Tolman-Oppenheimer-Volkov equations that yield sufficiently large
    maximal masses of neutron stars to be compatible with the recently
    observed PSR-J0952-0607 data as well as all other known radius and
    tidal deformation constraints.
  \end{abstract}
  \vfill
  \rule{10cm}{0.5pt}\\
  {\tt lorenzo@henu.edu.cn}, {\tt gudnason@henu.edu.cn}
\end{titlepage}

\pagenumbering{arabic}

\tableofcontents

\section{Introduction}

Neutron matter at densities present in the cores of neutron stars is
difficult to study for two reasons: first of all, nucleons at low
energy are described dominantly by the strong force, which is hard to
tackle due to the nonperturbativity of QCD at strong coupling.
The other reason making the study difficult, is that phenomenological
models that can capture nuclear physics at low densities are
practically impossible to extract reliably to large densities due to
the bottom-up approach of such models with a large number of effective
couplings that all, in principle, depend on the density.
Although perturbative QCD at extremely large densities is able to
predict reliably an equation of state \cite{Kurkela:2009gj}, such
densities are roughly an order of magnitude larger than what is needed for
neutron stars. 

Holographic QCD models, thanks to their ability to capture a large class of
nonperturbative phenomena, have recently been employed with some success
to provide predictions for nuclear matter at large densities, and hence represent
a valuable tool for the first-principles computation of the equation of state of nuclear matter 
in regimes typical for the core of neutron stars.
Holographic QCD models are a branch of AdS/CFT models developed after Maldacena
\cite{Maldacena:1997re} and Witten's work \cite{Witten:1998zw} which
are focused on the strong sector, viz.~Quantum Chromodynamics and they
come generally in two kinds: 
top-down models derived from string theory constructions, or bottom-up
models attempting to mimic the strong gluodynamics of QCD with an
appropriate 5-dimensional curved background.
In this work, we shall concentrate on the top-down model known as the
Witten-Sakai-Sugimoto (WSS) model \cite{Witten:1998zw,Sakai:2004cn,Sakai:2005yt},
which is especially predictive 
because it contains only two adjustable parameters, the Kaluza-Klein
scale ($M_{\rm KK}$) and the 't Hooft coupling $\lambda=g^2N_c$.
Usually these parameters are fitted to the meson sector of the theory,
reproducing the pion decay constant and the rho-meson mass
\cite{Sakai:2004cn}: when doing so, the model performs rather
poorly in the baryonic sector, while still providing some deep qualitative insights. 
To overcome this shortcoming, and to try to be as quantitatively precise as
the model's approximations allow, in this paper we shall employ a different
calibration of the model, which we will justify by the fact that we are working
exclusively with baryons in a homogeneous configuration, see below for
further discussion on the justification.

Holographic QCD models have been employed earlier in the literature in
order to describe neutron stars, in the WSS model
with taking the symmetry energy into account \cite{Kovensky:2021kzl},
in V-QCD \cite{Demircik:2022uol} which is a highly customized
phenomenological version of holographic QCD based on the Veneziano
limit -- i.e.~the limit of a large number of color as well as a large
number of quark flavors, in a D3/D7 model
\cite{Hoyos:2016zke,BitaghsirFadafan:2019ofb,BitaghsirFadafan:2020otb}, as 
well as in a hard-wall model \cite{Bartolini:2022rkl} (see Ref.~\cite{Jarvinen:2021jbd,Hoyos:2021uff} for nice reviews on the topic).
Baryons in the WSS model are instantons in the 4-dimensional subspace
spanned by the spatial and holographic directions of the 5-dimensional
curved AdS-like spacetime and are initially best understood via the
BPST flat-instanton approximation \cite{Hata:2007mb}, which is later
shown to be valid only in the large 't Hooft coupling limit
\cite{Bolognesi:2013nja}.
In the context of neutron stars it was shown that the instanton in the
pointlike approximation fails to describe tidal deformabilities and
baryon masses at the same time \cite{Zhang:2019tqd}.
Using instead the homogeneous Ansatz for nuclear matter has previously
been quite successful in describing neutron stars, at least barring
changing the meson fit of the WSS model to one more appropriate for
baryonic matter at finite densities \cite{Kovensky:2021kzl}.
Although the work \cite{Kovensky:2021kzl} holographically takes the
symmetry energy \cite{Kovensky:2021ddl}, charge neutrality 
and $\beta$-equilibrium into account and dynamically determines the
crust of the star, their approach has been plagued with
unrealistically large symmetry energies, a factor of 30 bigger than
its phenomenological value, which in turn shows up as large deviations
of physical neutron stars with respect to isospin symmetric neutron
stars, for example in the mass-radius data.
Such a large overestimation of the symmetry energy can be traced back
to the combination of two factors: on one hand one of the gauge fields
(the Abelian spatial component) is taken to vanish, an approximation
reliable in the large $\lambda$ limit, but not upon extrapolation to
finite $\lambda$, while another overall factor of $N_c^2$ compared to
our result arises from a different choice in the definition of the
large-$N_c$ isospin number for the nucleon states.

In a recent paper \cite{Bartolini:2022gdf}, the present authors have
employed a conceptually simpler approach to computing the symmetry
energy in the WSS model in the approximation of using the homogeneous Ansatz
and have shown that it can obtain realistic values, if the model is
calibrated with a smaller Kaluza-Klein scale and a larger 't Hooft
coupling, than the usual meson fit.
Not only is the WSS model able to reproduce the $\sim32$MeV symmetry
energy at saturation density, but the slope and second derivatives of
the symmetry energy as a function of the density can be made to be
consistent with all current experimental constraints -- coming both
from nuclear physics as well as from neutron star data.

In this paper, we use this result of a realistic symmetry energy as a
function of the density and to ensure this is the case, we choose to
bootstrap the model by setting the saturation energy to the physically
accepted one ($\sim0.16\fm^{-3}$), which fixes the Kaluza-Klein scale
and then adjust the 't Hooft coupling to obtain the correct value of
the symmetry energy at saturation density.
It turns out that this calibration scheme gives values of the
couplings very close to
those for which the symmetry energy passes all current experimental
bounds as a function of the density.
As the final ingredient in obtaining realistic neutron stars, requiring a
physical equation of state from very low to very high (medium)
densities, we take the hybrid approach in this paper as already
introduced in Ref.~\cite{Hoyos:2016zke}, by
patching/gluing a very reliable equation of state from nuclear physics
at low densities together with our equation of state obtained from the
WSS model in our calibration.

This paper is organized as follows.
In Sec.~\ref{sec:WSS} we introduce the WSS model as well as our
notation.
We review the static homogeneous Ansatz in Sec.~\ref{sec:static_homo},
and include time dependence in Sec.~\ref{sec:dynamic_homo}.
In Sec.~\ref{sec:beta_eq}, we review the relations for obtaining
charge neutrality, $\beta$-equilibrium and the corresponding
thermodynamic relations.
We then describe our fit and the hybrid equation of state in
Sec.~\ref{sec:EOS} and present the results of the neutron star
mass-radius diagrams in Sec.~\ref{sec:neutron_stars}.
We conclude the paper in Sec.~\ref{sec:conclusion} with a discussion
and outlook.

\section{Witten-Sakai-Sugimoto model}\label{sec:WSS}

The holographic model we choose to employ for the description of
nuclear matter at high density is the Witten-Sakai-Sugimoto model
\cite{Sakai:2004cn,Sakai:2005yt}, a top-down model of holographic QCD
based on the engineering of a couple of stacks of $N_f$
$D8/\overline{D8}$-branes on the supergravity background sourced by
$N_c$ $D4$-branes.
We will work in the configuration of antipodal branes and in the
confined geometry phase, so that the flavor branes will extend all the
way down to the tip of the cigar subspace, spanning all the
holographic direction.

In absence of a mass term for quarks (which we will neglect), the
theory after dimensional reduction to five dimensions is given by a
Dirac-Born-Infeld action supplemented by a Chern-Simons term.
Replacing the Dirac-Born-Infeld action at quadratic order in the field
strength, results in a Yang-Mills theory in curved space, with the
addition of the topological Chern-Simons term, so that the full action
within these approximations read
\beq
S=-\kappa \Tr\int\d^4x\d z\left[\frac{1}{2}h(z)\mathcal{F}_{\mu\nu}^2+k(z)\mathcal{F}_{\mu z}^2\right]+\frac{N_c}{24\pi^2}\int_{M_5}\omega_5,
\eeq
with the parameter $\kappa$ and the warp factors $k(z),h(z)$ given by
\beq
k(z)=1+z^2,\qquad
h(z)=\left(1+z^2\right)^{-\frac{1}{3}},\qquad
\kappa=\frac{\lambda N_c}{216\pi^3},
\eeq
where $\lambda$ is the 't Hooft coupling and $N_c$ is the number of
colors.
The $\U(2)$ gauge field 1-form $\calA$ is split into $\SU(2)$ and $\U(1)$
factors as
\beq
\calA = A^a \frac{\tau^a}{2} + \widehat{A}\frac{\mathbf{1}_2}{2},
\label{eq:notation}
\eeq
the field strength is given by $\calF=\d\calA+\calA\wedge\calA$,
$\tau^a$ are the standard Pauli spin matrices,
and the spacetime indices are denoted as
\beq
\alpha,\beta,\ldots = {0,M} \quad ;\quad M,N,\ldots = {i,z} \quad;\quad i,j,\ldots = {1,2,3}.
\eeq
The Chern-Simons 5-form $\omega_5$ is given by
\beq
\omega_5=\Tr\left(\mathcal{A}\wedge\mathcal{F}^2-\frac{\i}{2}\mathcal{A}^3\wedge\mathcal{F}-\frac{1}{10}\mathcal{A}^5\right),
\eeq
where the powers of forms are understood by the wedge product.
The $\U(N_f)$ gauge field describes flavor degrees of freedom: In the
present work we will assume the presence of two light flavors, thus
setting $N_f=2$ and neglecting contributions from the presence of the
strange quark, leaving their inclusion for future improvements. 
With the notation \eqref{eq:notation} we can rewrite the Chen-Simons term as
\beq
S_{\rm CS}=\frac{N_c}{24\pi^2}\int_{M_5}\Tr\left[3\widehat{A}\wedge F^2+\widehat{A}\wedge\widehat{F}^2+\d\left(\widehat{A}\wedge\left(2F\wedge A-\frac{\i}{2}A^3\right)\right)\right].
\eeq
We will assume the gauge fields $\widehat{A}_i,A_i$ to vanish at
$z=\pm \infty$, hence we drop the total derivative term. 
The top-down nature of the model enables it to only rely on two free
parameters to be fitted: the 't Hooft coupling $\lambda$ and the
Kaluza-Klein scale $M_{\rm KK}$. Note that the scale does not appear in
any of our expressions when doing calculations within the model: it is
the only mass scale of the model, and we can freely choose to work in
units of $M_{\rm KK}$, and then restore the correct power of $M_{\rm KK}$ to
obtain the results in physical units.

\section{Holographic nuclear matter: static homogeneous Ansatz}\label{sec:static_homo}

In the context of holographic QCD, baryons are described as
topological solitons of the flavor fields, and the
Witten-Sakai-Sugimoto model is no exception \cite{Bolognesi:2013nja}. 
The Witten-Sakai-Sugimoto soliton can be approximated as a BPST
instanton \cite{Hata:2007mb,Hashimoto:2008zw} in the limit of large $\lambda$: its
instanton number is then identified as the baryon number, allowing for
the description of nuclei on top of single baryons.

To exactly describe infinite nuclear matter in this model is to solve
a many-soliton problem in five-dimensional curved space time: this
task is too hard to be carried out for any practical application, so
what is done in most cases is to rely on some combination of
approximations and educated guesses. A possibility is to arrange
individual solitons in a lattice \cite{Baldino:2021uie}, minimizing
the free energy density to determine the elementary cell shape and
size (hence the density): this kind of approach is generally more
reliable at low density, when the individual baryons are well
separated, and approximations made to compute the interactions between
solitons are well under control. 

At densities around nuclear saturation and above, another possible
approximation exists: since nucleons are tightly packed, we can
approximate their spatial distribution in three-dimensional space to
be homogeneous, forming a uniform distribution of continuous nuclear
matter. 
In this process every information about individual baryons is lost, in
favor of intensive quantities such as the baryonic and isospin
densities.
Despite such a configuration not existing under assumptions of
homogeneity and regularity of the gauge fields as shown in
Ref.~\cite{Rozali:2007rx}, it turns out that it is still possible to
define a homogeneous Ansatz by modifying these assumptions, either by
enforcing homogeneity at the level of the field strengths
\cite{Elliot-Ripley:2016uwb}, or by introducing a discontinuity in the
homogeneous gauge fields \cite{Li:2015uea}.
For the purpose of this work, we will follow the second route,
introducing a discontinuity in the $\SU(2)$ gauge field that will
source a finite baryon density.
The great simplification that this  
approach produces lies in the reduction of complicated sets of PDEs to
more manageable sets of ODEs, the only remaining variable being the
holographic coordinate $z$ (and time, whose inclusion we will discuss
in the next section).  
The homogeneous Ansatz in the static approximation is given by the
gauge field configuration 
\begin{align}
  \calA^{\rm cl}_0 = \tfrac12\wha_0 \mathds{1},\quad
  \calA^{\rm cl}_i = -\tfrac12H\tau^i,\quad
  \calA^{\rm cl}_z = 0,
  \label{eq:HomStatic}
\end{align}
with $\wha_0=\wha_0(z), H=H(z)$ being functions of only the holographic coordinate $z$.

The function $H(z)$ encodes baryonic density, as it can be thought of
as the space-averaged many-soliton distribution. It does so in a
nontrivial way if we allow it to be discontinuous: for simplicity we
will assume a discontinuity is present in the function $H(z)$ at
$z=0$, which is the coordinate at which single solitons sit to
minimize energy, however it is possible to consider configurations
with more discontinuities located at finite $z=\pm z_0$ (see
Ref.~\cite{CruzRojas:2023ugm} for the treatment in both the
Witten-Sakai-Sugimoto and the VQCD models). 
The baryon density $d$ is given as 
\begin{align}
  d&=\frac{1}{32\pi^2}\int_{-\infty}^{+\infty}\d z\;\epsilon^{MNPQ}\Tr F_{MN}F_{PQ}\nonumber\\
  &=-\frac{3}{8\pi^2}\int_{-\infty}^{+\infty}\d z\; H'H^2 \nonumber \\
  &=-\frac{1}{8\pi^2}\left[H^3\right]^{z=+\infty}_{z=0^+}-\frac{1}{8\pi^2}\left[H^3\right]_{z=-\infty}^{z=0^-}.
\end{align}
We see immediately that a continuous function $H(z)$ would not be able
to describe nuclear matter at any finite density: it has to be an odd,
discontinuous function of $z$. We choose the function $H(z)$ to be
vanishing at the UV boundary $z=\pm \infty$, so that we are left with
an IR boundary condition for $H(z)$ reading 
\beq
H(0^\pm) = \pm\left(4\pi^2 d\right)^{\frac{1}{3}}.
\eeq
From now on, we will always perform integrals over $z$ on the "positive
$z$" half of the connected branes, accounting for the symmetry of the
integrands with an overall factor of two. Note that despite the
discontinuity in $H(z)$, the field strengths $F_{MN}^a$ are still
continuous, even functions of $z$ (though in general not
differentiable at $z=0$). 

The asymptotic UV value assumed by the function $\wha_0(z)$ is mapped
into the baryonic chemical potential as per the standard holographic
dictionary: we can then impose boundary conditions  
\begin{align}
  &\wha'_0(0) = 0,\qquad& \wha_0(\infty) &= \mu,\\
  &H(0)=\left(4\pi^2 d\right)^{\frac{1}{3}},\qquad& H(\infty)&=0,
\end{align}
and numerically solve the equations of motion 
\begin{align}
  h H^3 - \frac12\p_z(k H') - \frac{N_c}{32\pi^2\kappa}H^2\hat{a}_0' &= 0,\\
  \p_z(k\hat{a}_0') + \frac{3N_c}{16\pi^2}H^2H' &= 0.
\end{align}
With the normalization chosen for the asymptotics of $\wha_0$, the
baryon number chemical potential $\mu_B$ is given by 
\beq
\mu_B = \frac{N_c}{2}\mu.
\eeq
For every value of the parameter $\mu$, the corresponding
thermodynamic equilibrium value of the density $d(\mu)$ is given by
minimizing the free energy $\Omega$, holographically dual to the
on-shell action as $\Omega = - S^{\text{on-shell}}$. The baryonic phase is
favored over the vacuum when $\Omega(d,\mu)\leq 0$, a condition that
is satisfied for $d(\mu)\geq d_0(\mu_{\rm onset})$. The value $d_0$ is the
model prediction for nuclear saturation density of symmetric matter,
and will play an important role in our choice for the fit of the free
parameters $\lambda,M_{\rm KK}$.

\section{Isospin asymmetry: time-dependent homogeneous Ansatz}\label{sec:dynamic_homo}

The time-independent configuration introduced in the previous section
only accounts for symmetric baryonic matter. 
This can be understood by thinking about the single-baryon: a single
static soliton is a classical object, that carries no information
about the nucleon states. To have a spectrum including isospin states
we have to perform moduli space quantization \cite{Hata:2007mb}: an
arbitrary rotation in $\SU(2)$-space that corresponds to a zeromode is
introduced by performing the transformation $\tau^i\rightarrow a
\tau^i a^{-1}$. The orientation matrix $a$ is then promoted to a
time-dependent operator $a(t)$, in terms of which it is possible to
compute an effective quantum mechanical Hamiltonian that will give the
single-baryon spectrum. 
The same procedure can be applied to the homogeneous Ansatz, as
described in detail in Ref.~\cite{Bartolini:2022gdf}. In particular, the
homogeneous Ansatz is similar in its structure to the BPST instanton,
but simpler in that it lacks the position and size moduli. 
We then follow the same procedure in the treatment of rotational
moduli $a(t)\in\SU(2)$, and we define the angular velocity $\chi^i$
as: 
\beq
\chi^i \equiv -\i\Tr\left(a^{-1}\dot{a}\tau^i\right).
\eeq
The introduction of the $\SU(2)$ moduli and their time dependence turns
on new components of the gauge field, so that the new homogeneous
Ansatz for the iso-rotating configuration is given by 
\begin{align}
  \calA_0 = G a\bchi\cdot\btau a + \tfrac12\wha_0,\qquad
  \calA_i = -\tfrac12(H a \tau^i a+L\chi^i),\qquad
  \calA_z = 0,
  \label{eq:Ansatzchi}
\end{align} 
The classical action obtained from the time-dependent homogeneous
Ansatz following the prescription introduced in Ref.~\cite{Bartolini:2023eam} to 
choose the correct Chern-Simons term, is 
\begin{align}
  S_{\rm YM} = -\kappa\int\d^4x\int_0^\infty\d z\bigg[
    -&8hH^2\left(G+\frac12\right)^2\bchi\cdot\bchi
    +3hH^4\non
    &+k\left[(L')^2  - 4(G')^2 + 8(KH)^2\right]\bchi\cdot\bchi
    +3k(H')^2
    -k(\hat{a}_0')^2
    \bigg],\non
  S_{\rm CS} = -\frac{3N_c}{8\pi^2}\int\d^4x\int_0^\infty\d z&\hat{a}_0H'H^2\non
  +\frac{N_c}{4\pi^2}&\int\d^4x\int_0^\infty\d z
  \left(2LH'\left(G+\frac{1}{2}\right)+LHG'\right)H\bchi\cdot\bchi,
\end{align}
from which the equations of motion for the fields
$H(z),\wha_0(z),L(z),G(z)$ can be derived: 
\begin{align}
  h H^3 - \frac12\p_z(k H') - \frac{N_c}{32\pi^2\kappa}H^2\hat{a}_0' &= 0,\\
  \p_z(k\hat{a}_0') + \frac{3N_c}{16\pi^2}H^2H' &= 0,\\
  \p_z(k G') - 2h H^2\left(G + \frac12\right) + \frac{N_c}{32\pi^2\kappa} H^2L' &= 0,\\
  \p_z(k L') + \frac{N_c}{8\pi^2\kappa} H\left[H G' + (1 + 2G)H'\right] &=0.
\end{align}
The equations of motion have to be supplemented with adequate boundary conditions:
following Ref.~\cite{Bartolini:2023eam} we choose them so to cancel the infrared localized terms
in the variation of the action, which amounts to require:
\begin{align}
\left[G'(0)+\frac{N_c}{32\pi^2\kappa}H(0)^2L(0)\right]=0,\\
L'(0)=0.
\end{align}
The field $A_0$ encodes via the holographic dictionary the information on
the isospin density $n_I$ and chemical potential $\mu_I$. This can be
understood as follows: $G(z)$ can be thought of as a holographic profile for
the angular velocity $\chi^i$, which can be traded for the
corresponding angular momentum and then canonically quantized in the
moduli space approximation. The resulting operator is identified with
both spin and isospin operators (an artifact of the spherical symmetry
of the setup). 
When we do this, the resulting quantum Hamiltonian operator is given
by  
\beq	
H= \frac{I^2}{2V\Lambda} + VU,
\eeq
with $V$ being the three-dimensional (infinite) volume, $I^2$ is the
squared isospin operator, and $\Lambda$ and $U$  given by 
\begin{align}
  \Lambda &= 2\kappa
  \int_0^{\infty}\left[2h H^2(2G+1)^2 + k((L')^2 + 4(G')^2)\right]\d z,\non
  U &= \kappa
  \int_0^{\infty}\left[3h H^4 + 3k(H')^2 + k(\wha_0')^2\right]\d z.\label{eq:LambdaU}
\end{align}
They are respectively the moment of inertia of a rigid rotor and its
internal energy density, which in our physical setup corresponds to
the energy density of symmetric nuclear matter. 

From this Hamiltonian operator, we obtain the energy per nucleon of an
isospin state $|i,i_3\rangle$, from which we can read off the
symmetry-energy parameter of infinite nuclear matter $S_N(d)$: 
\beq
\frac{E}{A} =\frac{U}{d} + S_N(d)\beta^2
+ \mathcal{O}(V^{-1}),\qquad
S_N(d) = \frac{d}{8\Lambda},
\eeq
with $\beta=(N-Z)/A$ being the isospin asymmetry parameter.

We will also need the energy density since our final aim is to obtain
an equation of state for isospin asymmetric nuclear matter in
$\beta$-equilibrium with leptons. 
From the expressions above, the energy density of isospin-asymmetric
matter is obtained from $U,\Lambda$ as 
\begin{align}
  \mathcal{E}=\frac{E}{V}=&\frac{i_3^2}{2V^2\Lambda} + U
  = \frac{1}{2\Lambda}n_I^2 + U.\label{eq:epsilonnIU}
\end{align}
The isospin chemical potential $\mu_I$ can be obtained from the
expression above by differentiating with respect to $n_I$, so that  
\beq\label{eq:muI}
\mu_I = \frac{\partial\mathcal{E} }{\partial n_I} = \frac{n_I}{\Lambda}.
\eeq
Note that we have defined $n_I$ from the isospin quantum number $i_3$,
which assume values  $\pm 1/2$ for the ground states. We choose to
identify the positive eigenvalue to correspond to the proton, and the
negative one to the neutron, following conventions from nuclear
physics. 
With this choices, if the nuclear matter is rich in neutrons (as we
expect it to be following beta equilibrium and charge neutrality), we
will have $n_I < 0$ and correspondingly also the isospin chemical
potential $\mu_I$ will be negative due to Eq.~\eqref{eq:muI}.

\section{\texorpdfstring{$\beta$}{beta}-equilibrated matter}\label{sec:beta_eq}

The presence of the symmetry energy favors isospin-symmetric matter:
the neutron rich matter that composes the core neutron stars is the
result of the presence of negatively charged leptons that impose
electric charge neutrality to the system. The leptons with their
(anti)neutrinos, the neutrons and the protons are in equilibrium with
respect to $\beta$-decay. This condition, together with charge
neutrality, is sufficient to calculate the fractions of each particle
species at any given density. 
As discussed in Ref.~\cite{Bartolini:2022gdf} with the same normalizations
used in this work, the imposition of $\beta$-equilibrium (together
with the decoupling of the neutrinos) and electric charge neutrality
lead respectively to the conditions 
\beq
\mu_\ell &=& \mu_N - \mu_P = -\mu_I, \qquad
\ell=e,\mu, 
\label{eq:beta_equilibrium}\\
\frac{1}{2} d +n_I &=& {\textstyle\sum}_\ell n_\ell,
\label{eq:charge_neutral}
\eeq
where $\mu_X$ is the chemical potential of the particle species $X$
and we accounted for the presence of electrons ($e$) and muons ($\mu$)
in equilibrium with protons ($P$) and neutrons ($N$), while neglecting
the presence of the heavier taus ($\tau$). 

The leptonic number densities $n_\ell$ are taken as originating from the
free energy of a (massive) Fermi gas: 
\begin{equation}\label{eq:Omegalepton}
\Omega_\ell = -\frac{1}{24\pi^2}\Theta\left(\mu_\ell-m_\ell\right)\left[\left(2\mu_\ell^2-5m_\ell^2\right)\mu_\ell\sqrt{\mu_\ell^2-m_\ell^2}+3m_\ell^4\log\left(\frac{\sqrt{\mu_\ell^4-m_\ell^4}+\mu_\ell}{m_\ell}\right)\right],
\end{equation}
from which follows
\beq
n_\ell(\mu_\ell)=\Theta_H(\mu_\ell-m_\ell)\frac{(\mu_\ell^2-m_\ell^2)^{\frac32}}{3\pi^2}.
\label{eq:Fermi_rho}
\eeq
We will approximate the electrons to be massless, while keeping the
muons massive with $m_\mu^{\rm phys}=105.66$MeV, while the parameter
appearing in the equations is the corresponding value in units of
$M_{\rm KK}$, $m_\mu = m_\mu^{\rm phys} M_{\rm KK}^{-1}$.\footnote{Note
that by doing this we are introducing another energy scale in the
model: since the value of $m_\mu$ depends on the choice of the value
$M_{\rm KK}$, such parameter can no longer be factorized out from
the equations.}

Inserting Eq.~\eqref{eq:Fermi_rho} into the charge-neutrality
condition \eqref{eq:charge_neutral} and using the $\beta$-equilibrium
condition \eqref{eq:beta_equilibrium}, we obtain an implicit solution
for the isospin density, $n_I$, as a function of the baryon density $d$ 
(note that with the present conventions, $n_I$ is always negative
in neutron rich matter, and that $\Lambda$ is always positive by construction,
as appropriate for its interpretation as a moment of inertia):
\begin{align}
  \frac{n_I^3}{3\pi^2\Lambda^3}
  \left[\Theta_H(-n_I)
    +(1-R^{-2}m_\mu^2)^{\frac32}\Theta_H(-R-m_\mu)
    \right] 	\mathop+n_I + \frac12 d = 0, \qquad R\equiv n_I/\Lambda.
\end{align}
From $n_I(d)$ it is then possible to compute the density of every
particle species as 
\begin{align}
  n_P(d)&=\frac{1}{2}d + n_I(d), \label{eq:populationP}\\
  n_N(d)&=\frac{1}{2}d - n_I(d),\label{eq:populationN}\\
  n_e(d)&=\frac{1}{3\pi^2}\left(\mu_I(d)^2\right)^{\frac{3}{2}},\label{eq:populatione}\\
  n_\mu(d)&=\frac{1}{3\pi^2}\left(\mu_I(d)^2-m_\mu^2\right)^{\frac{3}{2}}.\label{eq:populationmu}
\end{align}

To build an equation of state for the matter we just described, we
need now to compute the total energy density and pressure. The
building blocks are the energy densities $U$ and pressure $P_0$ of
symmetric nuclear matter, the contributions $\mathcal{E}_I$ and $P_I$
due to isospin asymmetry, and the contributions $\mathcal{E}_\ell,
P_\ell$ of leptons. 

We already know an expression for $U$ from Eq.~\eqref{eq:LambdaU}, which
we can use to compute the pressure $P_0$ using $\mu_B$ and $d$ 
\beq
P_0 = -U + \mu_B d.
\eeq
For the contributions arising from isospin asymmetry, we make use of 
the known relations \eqref{eq:muI}, \eqref{eq:epsilonnIU}:
\beq
\mathcal{E}_{I}&=& \frac{1}{2\Lambda} n_I^2,\\
P_{I}&=& -\mathcal{E}_I + \mu_I n_I = \frac{1}{2\Lambda}n_I^2 = \mathcal{E}_{I}.
\eeq
Finally, for the leptons we can use the fact that in a homogeneous
system $P=-\Omega$ holds, so that we can compute $P_\ell$ directly
from Eq.~\eqref{eq:Omegalepton}, from which the energy density follows
\begin{align}
  \mathcal{E}_e+	\mathcal{E}_\mu &= \Omega_e+\Omega_\mu +\mu_e n_e + \mu_\mu n_\mu\non
  &=  \Omega_e+\Omega_\mu -\mu_I n_P\non
  &= \Omega_e+\Omega_\mu -\frac{n_I}{\Lambda}\left( \frac{1}{2}d+n_I\right),
\end{align}
where we have made use of charge neutrality and $\beta$-equilibrium.
As a result, we can finally write the total energy density and
pressure as 
\beq
\mathcal{E}&=& U+\mathcal{E}_I+\mathcal{E}_e+\mathcal{E}_\mu,\\
P &=& P_0 + P_I-\Omega_e - \Omega_\mu,
\eeq
with every quantity involved being a function of only the baryon
density $d$. 

\section{Nuclear matter fit and hybrid equation of state}\label{sec:EOS}

\begin{figure}
  \centering
  \includegraphics[width=0.7\linewidth]{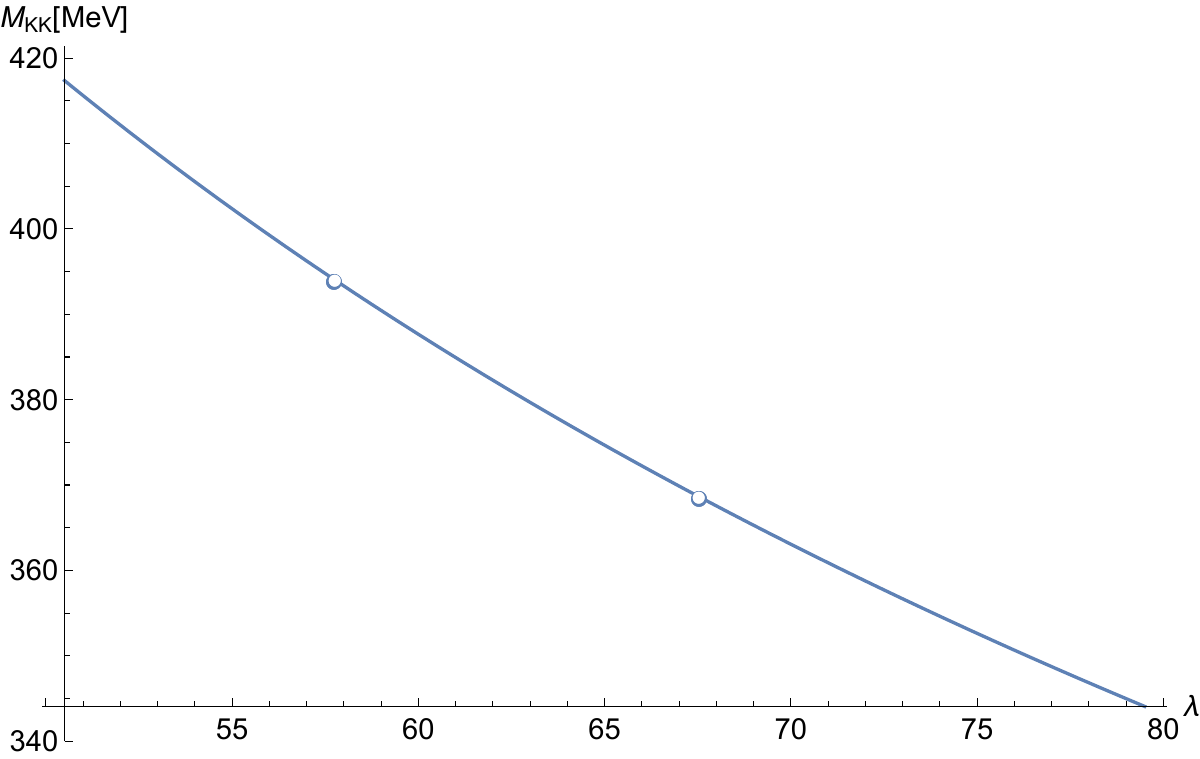}
  \caption{The curve $M_{\rm KK}(\lambda)$ that fits the saturation
    density of symmetric nuclear matter to its phenomenological value
    of $d_0^{\rm phys}=0.16\fm^{-3}$. The two dots marked on the plot
    correspond to the two fits we employed, corresponding to the upper
    and lower bounds on the symmetry energy of $32.8\MeV$ and
    $30.6\MeV$, respectively.}
  \label{fig:MkkLambda} 
\end{figure}
It is well known that the most common fit of the Witten-Sakai-Sugimoto
model ($M_{\rm KK}=949$MeV, $\lambda=16.6$) performs very well in the
mesonic sector, but that its quantitative reliability when employed to
compute quantities of the baryonic sector is somewhat lacking (in
particular the masses of the baryons are largely overestimated,
together with the binding energies of nuclei). 
When we move to the dense nuclear matter described by the homogeneous
Ansatz, the situation gets worse and contact with phenomenology is
lost. For example, the saturation density for $\lambda = 16.6$ is
found to be $d_0^{\lambda=16.6} = 0.00385 M_{\rm KK}^{3}$, while the
symmetry energy at saturation is of order
$S_N^{\lambda=16.6}(d_0^{\lambda=16.6})\simeq 0.1 M_{\rm KK}$: when
the value $M_{\rm KK}=949$MeV is plugged into these expressions, the
results are respectively $d_0=0.43\fm^{-3}$ and $S_N(d_0)=97$MeV.  
If we ignore these issues and try to build the core of neutron stars
with this fit, we unsurprisingly find that they never reach
$1.5$M$_{\odot}$, and their radius can at most be slightly larger than
$7$km.  
Realistic values for the highest mass for a stable star and for the
average radii have been obtained with a fully holographic equation of
state (though with the phenomenological input of the surface tension
of the domain wall between nuclear matter and vacuum) from the
Witten-Sakai-Sugimoto model in Ref.~\cite{Kovensky:2021kzl}: to obtain a
realistic-looking neutron star while keeping $M_{\rm KK}=949$MeV, it
is necessary to decrease $\lambda$ to around the value $\lambda=10$, with
the results being highly sensitive to changes around this
value. Realistic masses and radii for neutron stars as functions of the model
parameters $\lambda,M_{KK}$ within this setup have been further explored
in Ref.~\cite{Kovensky:2021wzu}. However, even producing stars with adequate radii
and masses, nuclear matter inside the stars of Ref.~\cite{Kovensky:2021kzl} 
fails to be as rich in neutrons as we expect in neutron stars:
this is due to the
extremely high value of the symmetry energy, which pushes matter
towards isospin symmetric configurations, and remains true even if we adopt
the lower values of symmetry energies found in Ref.~\cite{Bartolini:2022gdf}, as
long as we adopt fits defined in the ``QCD window'' introduced in Ref.~\cite{Kovensky:2021wzu}.
Our aim is to describe realistic neutron star cores and connect the
equation of state from holography with a phenomenologically accurate
equation of state at lower energy densities: within this approach, it
makes little sense to employ a fit that overestimates physical
quantities that are crucial in the formation of the structure of
neutron stars by a factor of roughly three. Moreover, the fit is done
on single-particle properties, but information on individual particles
is lost when employing the homogeneous Ansatz, which does not connect
smoothly to the finite particle number setup. 
\begin{figure}
  \centering
  \includegraphics[width=0.49\linewidth]{{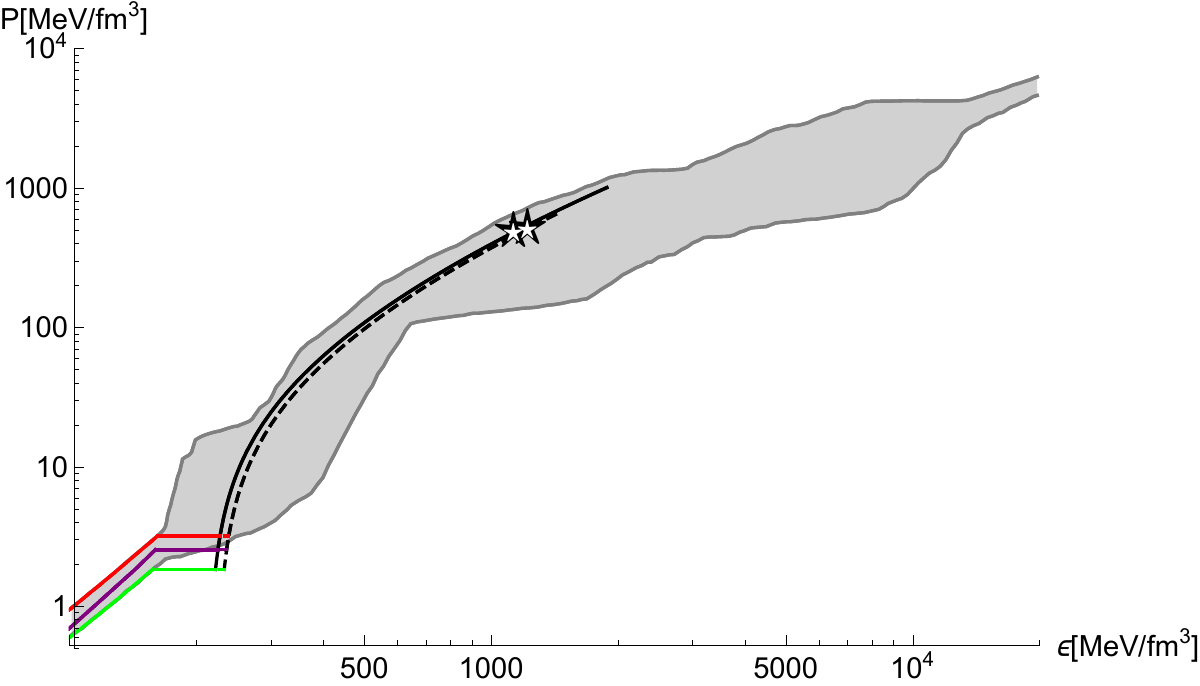}}  
  \includegraphics[width=0.49\linewidth]{{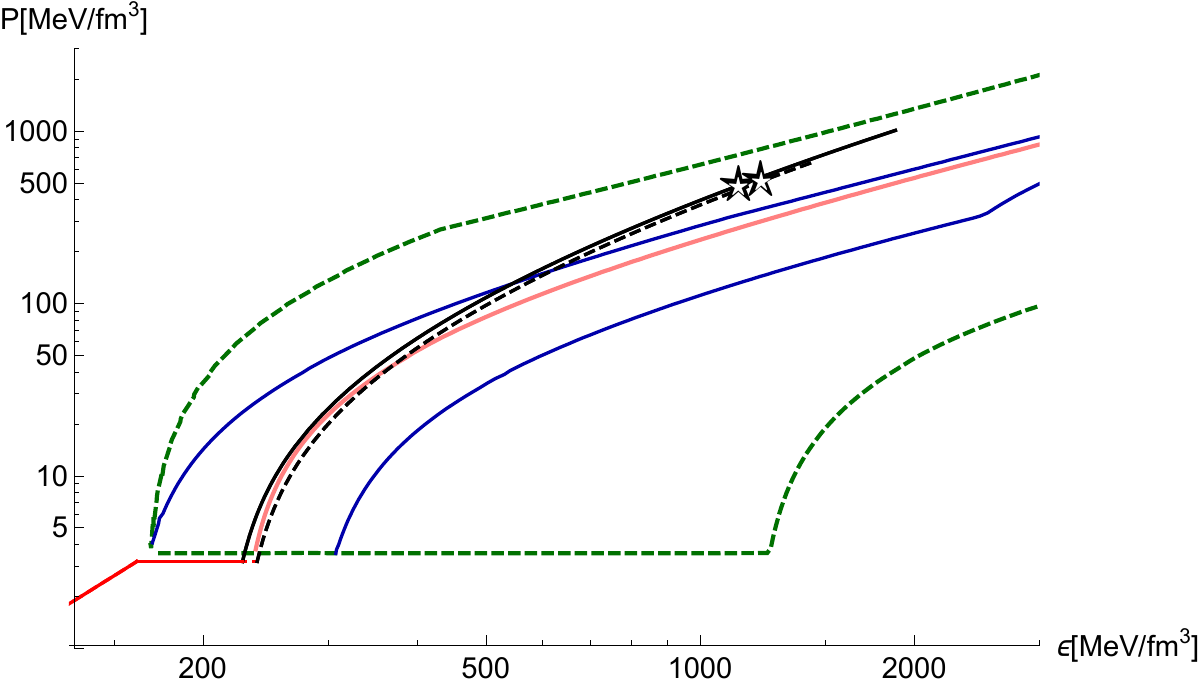}}
  \caption{Left panel: in red (green) the low-density equation of state 
      obtained from the upper (lower) bound derived from CEFT,
      in purple the low-density SLy4 equation of state,
      in black the branches of the equations of
      state computed from the WSS model, the gray-shaded area marks 
      currently accepted bounds for the EOS
      of nuclear matter constrained by neutron stars data. The solid (dashed) line
      represents the fit that achieves $S_N=32.8\MeV$ ($S_N=30.6\MeV$) at
      saturation density $d_0=0.16\fm^{-3}$. The star markers in
      the plot represent the pressure and density of the heaviest stable
      static neutron stars allowed by that particular equation of
      state (the left-most star marker corresponds to the solid line).
      Right panel: the hybrid stiff-WSS EOS compared with model independent
      bounds from Ref.~\cite{Komoltsev:2021jzg}, derived from pQCD and
      shown with a dashed green line.
      EOS lying within the dark blue lines
      have an upper-bound on the speed of sound at the conformal value $c_s^2<1/3$. 
      The pink line represents the limiting case in which the speed of sound is 
      constant along the EOS.
  }\label{fig:EOS} 
\end{figure}

We hence propose another choice of fit for the free parameters of the
Witten-Sakai-Sugimoto model, expected to give precise results when
applied to highly dense matter, especially in the context of the
homogeneous Ansatz. The first quantity we fit is the nuclear
saturation density for symmetric matter: phenomenologically this value
lies at about $0.16\fm^{-3}$. The value of the model-derived
saturation density in units of $M_{\rm KK}$ does not depend on $M_{\rm KK}$
itself, so we only need to compute the value $d_0(\lambda)$ for
a wide range of values of $\lambda$, then impose the condition 
\beq
d_0(\lambda) M_{\rm KK}^{3} =  0.16\fm^{-3}.
\eeq 
This defines the function $M_{\rm KK}(\lambda)$ we plot in figure
\ref{fig:MkkLambda}. The second choice we employ to eliminate any
further freedom is to have a realistic symmetry energy at saturation
density. The phenomenologically accepted values range between
$30.6\MeV$ and $32.8\MeV$: we fit the model to both values, so to
produce an "error bar" for our predictions. 
As a result, the two fits for which we will present results correspond
to 
\beq\label{eq:fitHlambda}	
 \lambda &=& 67.55\phantom{7}\qquad,\qquad M_{\rm KK} = 368.6\MeV\quad \Rightarrow \quad S_N(d_0) = 30.6 \MeV,\\
\label{eq:fitLlambda}
\lambda &=&  57.76\qquad,\qquad M_{\rm KK} = 394.01\MeV\quad \Rightarrow \quad S_N(d_0) = 32.8\MeV.
\eeq
Using these fits we build the equation of state for holographic
matter, which we assume to be the appropriate description at densities
above saturation: to describe lower-density regions of the neutron
star, namely the crust, we would need a more refined construction,
possibly deviating from the homogeneous approximation, for example
nuclear pasta phases. A similar construction was performed in
Ref.~\cite{Kovensky:2021kzl}, where a crust was built from the mixed phase
of lumps of nuclear matter in $\beta$-equilibrium, immersed in a gas
of leptons, with the surface tension of the domain wall, separating the
phases, as the parameter to regulate the onset of the transition from
homogeneous matter to the mixed phase. 
Here we follow another possible approach, that of a ``hybrid'' equation
of state: within this framework, we forego the attempt of building the
full equation of state from a single model, and instead patch together
equations of state coming from different models, each derived within
the domain of applicability of the models themselves.  
In particular, equations of state for low-density nuclear matter are
already developed with a higher precision than the one we can hope to
achieve within the holographic model: we then choose to employ 
a set of three different low energy equations of state: a soft and a stiff one, 
taken as the two limiting scenarios from chiral EFT interactions as described
in Ref.~\cite{Hebeler:2013nza}, and the 
SLy4 equation of state tabulated in Ref.~\cite{Douchin:2001sv,Haensel:1993zw}, which we
then match
at larger densities with the holographic equation of state
described in the previous section: the criteria we require for the matching
are that the pressure and baryon number density are continuous at the junction.

The density at which the patching is performed is weakly dependent on 
the fit choice and on the low-density equation of state employed, in the range
of $d_P\in(1.032 d_0, 1.059 d_0)$, with lower values for the soft equation of state, intermediate values 
around $1.045d_0$ for the SLy4, and the upper bound for the stiff equation of state.

By doing this, we obtain the equations of state shown in
Fig.~\ref{fig:EOS}, where they are compared with currently established 
bounds.

\begin{figure}
  \centering
  \includegraphics[width=0.7\linewidth]{{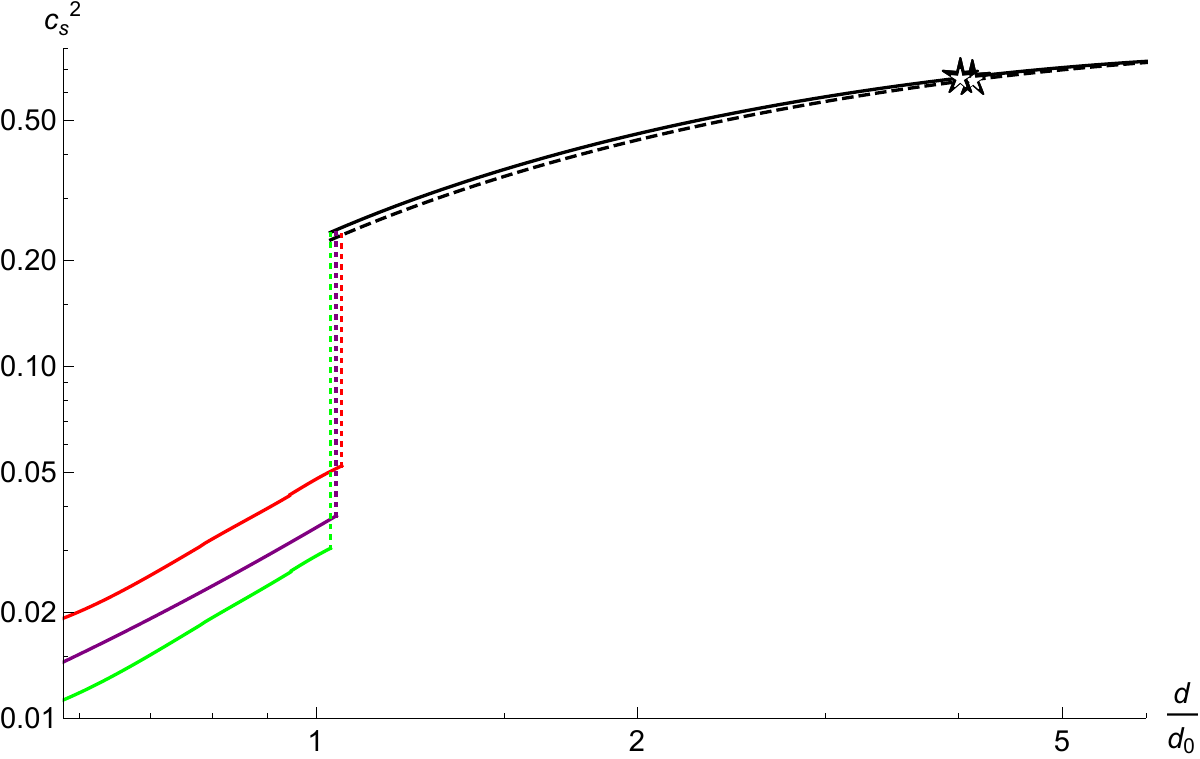}}
  \caption{The speed of sound for the three hybrid equations 
  of state presented in Fig.~\ref{fig:EOS} as function of density in 
  units of the saturation density $d_0$. The star markers represent the speed
  of sound reached inside the heaviest stable static neutron stars allowed by the corresponding 
  equation of state.}
  \label{fig:Speed}
\end{figure}

From the equations of state, we can numerically compute the speed of
sound: the resulting plot is shown in Fig.~\ref{fig:Speed}, and it
shows the presence of a discontinuity at the density corresponding to
the junction between the two equations of state from different
models. The holographic equation of state is highly stiff as it rapidly
crosses the conformal value of $c_s^2=1/3$: this is a feature it
shares with many equations of state derived from holography, but even
for this category it achieves very high sound speeds, with the
highest speed present in stable neutron stars ranging in the interval
$c_s^2\in\left(0.652,0.660\right)$, depending on the choice of fit (the
two star markers in Fig.~\ref{fig:Speed} pinpoint these
values). Note also that the stability of the stars is what limits the
speed of sound to these values: the equation of state extrapolated to
higher densities produces higher velocities, which however are never reached in stable stars.

\begin{figure}
  \includegraphics[width=0.49\linewidth]{{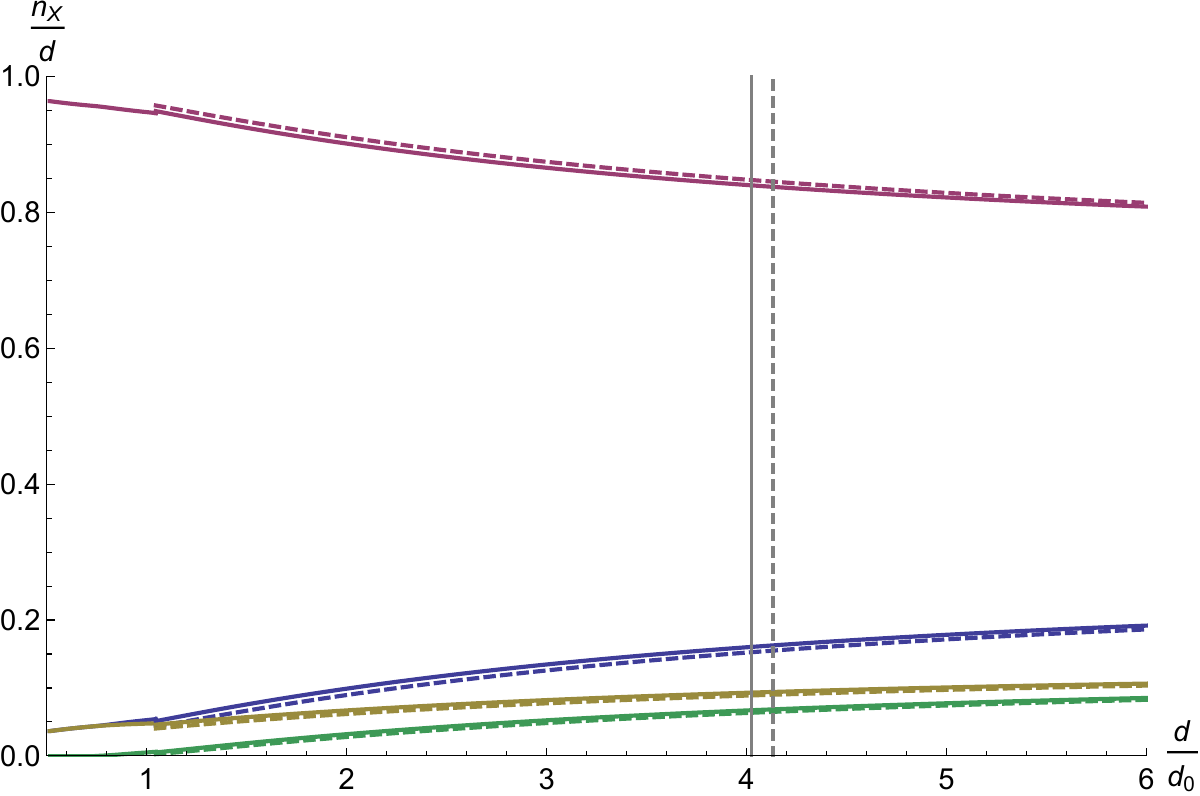}}
  \includegraphics[width=0.49\linewidth]{{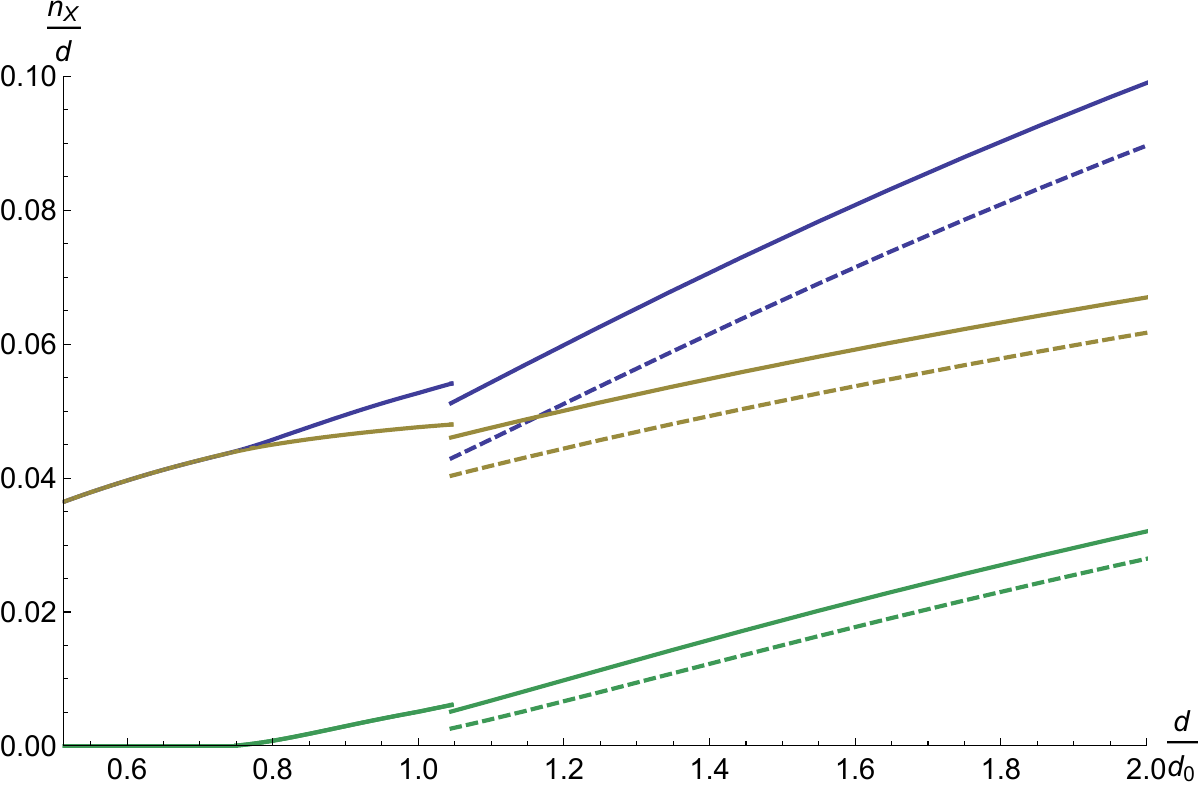}}
  \caption{Particle fractions as function of the density (in units of
    the saturation density), with the curves from top to bottom
    corresponding to neutrons (dark red), protons (purple), electrons
    (yellow) and muons (green).
    The solid and dashed lines are for the two different fits, see
    the caption of Fig.~\ref{fig:EOS}. 
    The left panel covers the entire range
    relevant for even our heaviest neutron stars, whereas the right
    panel zooms in around the densities where we have patched
    the SLy4 and the WSS equation of state together.}
  \label{fig:Fracs}
\end{figure}
We can also test our choices of fit against the particle populations:
using Eqs.~\eqref{eq:populationP} to \eqref{eq:populationmu} we
can compute the number densities of each species of particles, and
compare them with those of the SLy4 equation of state at the junction
point. 
The results are presented in the two panels of Fig.~\ref{fig:Fracs}:
we can see from the right panel that, while at the junction point
the SLy4 populations to not fall precisely within the predictions of our
two fit choices, the mismatch is quite small, and the results from the 
holographic model are well in the correct ballpark. From this result alone
it would seem that a favored fit would lie closer to that of Eq.~\eqref{eq:fitLlambda}
corresponding to $S_N(d_0)=32.8\MeV$, a tendency that will be reinforced
in the next section by comparing the results for the tidal deformabilities of 
neutron stars.

Before applying the computed equations of state to neutron stars, let us
comment on the shortcomings of the approach and approximations we employed.
The first noticeable feature shared by the equations of state plotted in Fig.~\ref{fig:EOS}
is the discontinuity in the energy density $\mathcal{E}$, represented by the horizontal
segment that connects the low-density branches to the holographic part: we regard this
unexpected phase transition as an artifact due to the failure of the 
model, the Ansatz and the fit choice, to reproduce the correct onset value of the baryonic chemical
potential, giving the value of $1345.5\MeV$ ($1409.9\MeV$) for $\lambda=57.76$ ($\lambda=67.55$).
A more sophisticated construction of the transition between the two equations of
states could in principle ameliorate the problem: at lower densities, just around
saturation, we expect the homogeneous approximation to lose reliability, while configurations
where baryons can individually be resolved become more realistic (e.g.~lattice configurations).
Connected to this issue, is the sharp discontinuity in the speed of sound, as manifest in Fig.~\ref{fig:Speed}.
In the same spirit, we should not regard the transition between the two equations of state as 
being exhaustively described by our construction since at the patching density $d_P$ the properties
of nuclear matter abruptly change behavior to that of the high-density regime:
again, we can expect that introducing at least one intermediate description could provide a more
realistic speed of sound around these densities. As an example, in Ref.~\cite{BitaghsirFadafan:2018uzs} nuclear matter
built from an instanton Ansatz leads to a speed of sound at saturation density of roughly $c_s^2(d_0)\simeq 0.025$,
a value close to the ones reached by the three phenomenological equations of state we employed at low density: 
this can be indicative that the instantonic description does indeed capture more reliably the physics of 
nuclear matter in this regime, though we have to keep in mind that the
brane configuration in Ref.~\cite{BitaghsirFadafan:2018uzs} is
different from the one we have employed.
The idea that the homogeneous approximation is to be held responsible for this sharp 
jump in the speed of sound is also strengthened by the observation that a similar behavior is 
shared between different constructions that employ the same description: in particular, it is 
present in Ref.~\cite{Kovensky:2021kzl}, which still uses the WSS model, but contains a holographic EOS 
also for the crust of the star, and it is present in VQCD as can be noted in the review \cite{Jarvinen:2021jbd}.

Lastly, we should comment on the high-density regime: we argued that the homogeneous Ansatz
becomes increasingly reliable as the density increases. While this is intuitively true, it still does
not take into account other possible effects that can lead to different physics. In particular, we need to 
address two shortcomings of our approach that manifest themselves
in the high-density limit. 

The first one is that in the present work we have not included the possibility of a phase transition to quark matter: as pointed out in Ref.~\cite{Annala:2023cwx}, it is expected that matter in the heaviest 
neutron stars exhibits a deconfined phase, and this behavior is reflected in the speed of sound approaching 
values close to the conformal limit $c_{s}^2=1/3$ already around the TOV densities. Our choice of not
exploring the possibility of this phase is mostly due to practical reasons: in the WSS model with antipodal
flavor branes, the inclusion of quark matter is only possible in the deconfined geometry (i.e.~the black brane horizon 
sources the quark baryonic charge), which requires (in the 
absence of backreaction of the flavor branes on the geometry) high
temperatures. In principle, including backreaction
could lead to such a phase even in the low-temperature regime, since a
high baryon density can also induce the formation of a horizon in the
bulk, which in turn would source a non-topological baryon number, to be identified as 
generated by individual quarks. However, the full backreaction of Witten's background is extremely challenging,
and its development is far from achieved (though the first order in $\frac{N_f}{N_c}$ has been developed, at least in 
the ``smeared'' configuration of the flavor branes, see Ref.~\cite{Bigazzi:2014qsa}). Another possibility would be to 
abandon the antipodal branes configuration, and then adopt the decompactified limit as in Ref.~\cite{BitaghsirFadafan:2018uzs,Kovensky:2019bih,Kovensky:2020xif}: in this way, at the price of
changing the branes' UV boundary condition (hence the dual UV theory) and introducing the dynamics of the 
embedding of the flavor branes to the equations of motion, it is possible to describe quark matter in the low-temperature
regime \cite{Kovensky:2019bih}, and possibly even quarkyonic matter \cite{Kovensky:2020xif}. We are leaving
the investigation of these possibilities to future developments.

The second limitation we need to point out is that, even if we
include the high-density regime, we cannot expect it to converge to
the correct asymptotic behavior of perturbative QCD: this is a
fundamental shortcoming of the model, that fails at reproducing
asymptotic freedom. As a result, even including a phase transition at
high density, there is little hope that the equation of state from
Fig.~\ref{fig:EOS} connects with the pQCD limit.
Despite this limitation, we stress that the part we computed of the
equation of state, does not violate the constraints set by 
Ref.~\cite{Komoltsev:2021jzg}, as shown in the right panel of
Fig.~\ref{fig:EOS}.  
To establish the comparison, we plot the same bounds as in
Ref.~\cite{Komoltsev:2021jzg}, together with our hybrid stiff-WSS EOS, as
appropriate for the bounds (chosen in the latter reference), obtained by matching the high-energy pQCD
EOS with the low-energy CEFT stiff EOS at $d=1.1d_0$. 
Our EOS is found to only lay below the lower bound in a very small
range of values around the hybridization density: this however is
simply due to the choice performed in Ref.~\cite{Komoltsev:2021jzg} to take
the low-density limit to be identified with $d=1.1d_0$, and
interpolate between that particular value and the high-density one
from pQCD, while for us the critical density of hybridization is
around $d_P^{\rm stiff}=1.06 d_0$. Choosing our $d_P^{\rm stiff}$ as a low
density limit would result in new bounds which establish the hybrid
EOS to be consistent with pQCD bounds at least up to TOV densities. 
However, as already pointed out, we expect the WSS model to exhibit
corrections to the computed behavior around saturation density, so we
leave a complete analysis around these regions to future works, when a
more quantitatively trustable approach will be available.

\section{Neutron stars}\label{sec:neutron_stars}

We finally want to derive properties of neutron stars built with the
equation of state we just constructed: to do so, we solve numerically
the Tolman-Oppenheimer-Volkov (TOV) equations for a range of central
pressures ($P_0$). Each solution will provide a neutron star with
central density $P_0$, radius $R(P_0)$, mass $M(P_0)$ and tidal
deformability $\Lambda(P_0)$. 
The TOV equations are given by 
\begin{align}\label{eq:TOV1}
  \frac{\d P}{\d r}&= -G(\calE+ P)\frac{m+4\pi r^3 P}{r(r-2Gm)},\\
  \frac{\d m}{\d r}&= 4\pi r^2 \calE,\label{eq:TOV2}
\end{align}
and describe static neutron stars, hence we are neglecting the effects
of rotation in this analysis. 
We supplement the equations with the third one
\begin{align}
  r\frac{\d y}{\d r}&=-y^2-\frac{4\pi G r^2\left(5\calE +9P+\frac{\calE+P}{c_s^2}\right)-6}{1-\frac{2GM}{r}}-\frac{y\left[1-4\pi Gr^2\left(\calE-P\right)\right]}{1-\frac{2GM}{r}}\non
&\phantom{=\ }
  +\frac{4G^2\left(M+4\pi Pr^3\right)^2}{r^2\left(1-\frac{2GM}{r}\right)^2}.
\label{eq:tidaldiff}
\end{align}

\begin{figure}
  \centering
  \includegraphics[width=0.7\linewidth]{{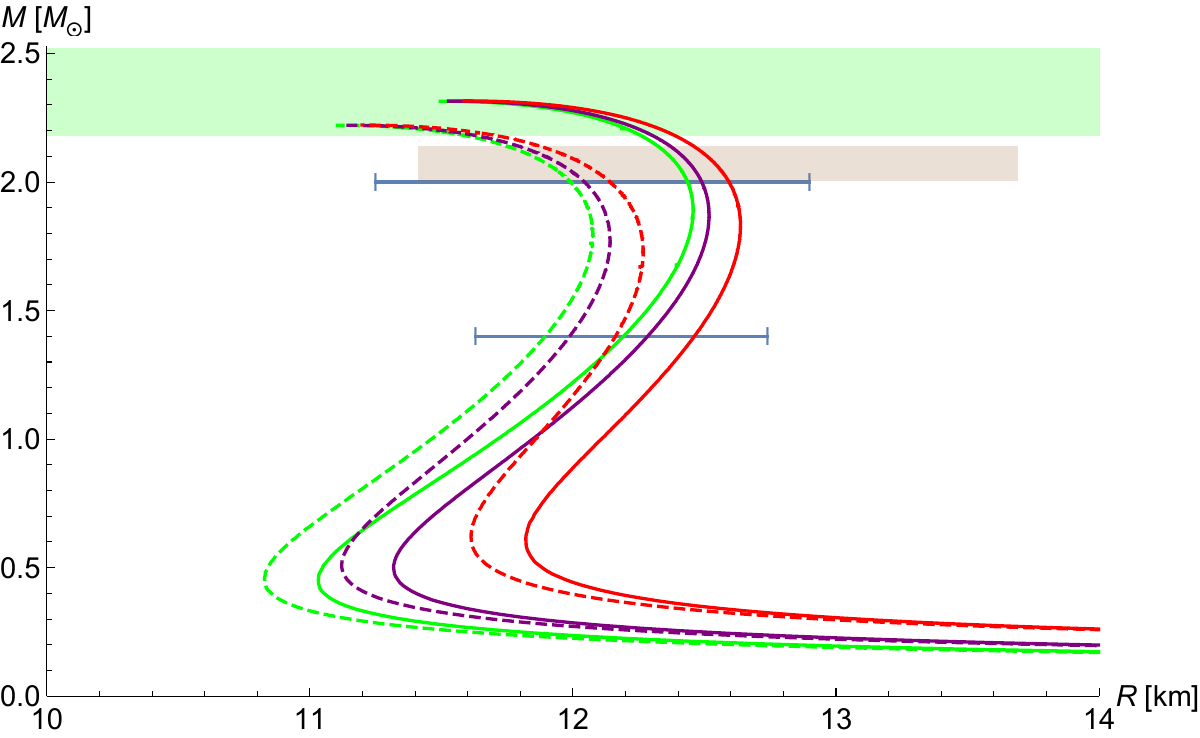}}
  \caption{The mass-radius diagram of neutron stars in our model. 
    In red (green), the curves corresponding to neutron stars built from
    the WSS equation of state hybridized with the stiff (soft) CEFT one. 
    In purple, the curves corresponding to neutron stars built from the 
    WSS equation of state hybridized with SLy4.
    For every color, the solid (dashed) line
    represent the fit that achieves $S_N=32.8\MeV$ ($S_N=30.6\MeV$) at
    saturation density $d_0=0.16\fm^{-3}$. The blue horizontal error bars
    are calculated from $208$Pb neutron skin thickness
    \cite{Lim:2022fap}, while the gray-shaded region is from
    Ref.~\cite{Riley:2021pdl}. The light green shaded area represents the
    measured mass of PSR J0952-0607 and represents a lower bound for
    the maximal achievable mass by a stable (static) neutron star.}
  \label{fig:MR} 
\end{figure}
From the solution of Eqs.~\eqref{eq:TOV1}-\eqref{eq:TOV2} we produce
relations between mass and radius, and employing Eq.~\eqref{eq:tidaldiff} 
with the boundary condition $y(0)=2$, we obtain the relation between mass and 
tidal deformability $\Lambda$, with the definition
\beq
\Lambda=\frac{2k_2}{3c^2}.
\eeq
In the above equation, $c=\frac{GM}{R}$ is the compactness of the star, while $k_2$ is the
tidal Love number, defined by:
\begin{align}
k_2=&\,\frac{8c^5}{5}\left(1-2c\right)^2\left[2-y_R+2c(y_R-1)\right]\times \left\{2c\left[ 6-3y_R+3c\left(5y_R-8\right) \right]+\right.\nonumber\\
&+\left. 4c^3\left[13-11y_R+c(3y_R-2)+2c^2(1+y_R)\right]\right.\nonumber\\
&\left.+3(1-2c)^2 \left[2-y_R+2c(y_R-1)\right]\ln\left(1-2c\right) \right\}^{-1}.
\end{align}
In particular the mass-radius curve represents the best test currently
available for neutron stars models: our results for this relation are
provided in Fig.~\ref{fig:MR}, where we plot in purple the data for the
stars generated by the SLy4-WSS equation of state, and in red (green) the 
data for the ones generated by the stiff (soft) CEFT-WSS equation.
Remarkably, all the hybrid equations of state, derived from fitting the WSS model
to saturation density and symmetry energy, result in neutron stars satisfying every
constraint in the mass-radius curve, and achieve rather high maximum
masses for stable stars, compatible with the highest mass measured to
date, i.e.~$2.35\pm0.17M_{\odot}$ of PSR J0952-0607\footnote{Though
this constraint does not necessarily need to be satisfied by static objects since
PSR J0952-0607 is a pulsar with a $1.41$ms rotation period.}.

\begin{figure}
  \centering
  \includegraphics[width=0.7\linewidth]{{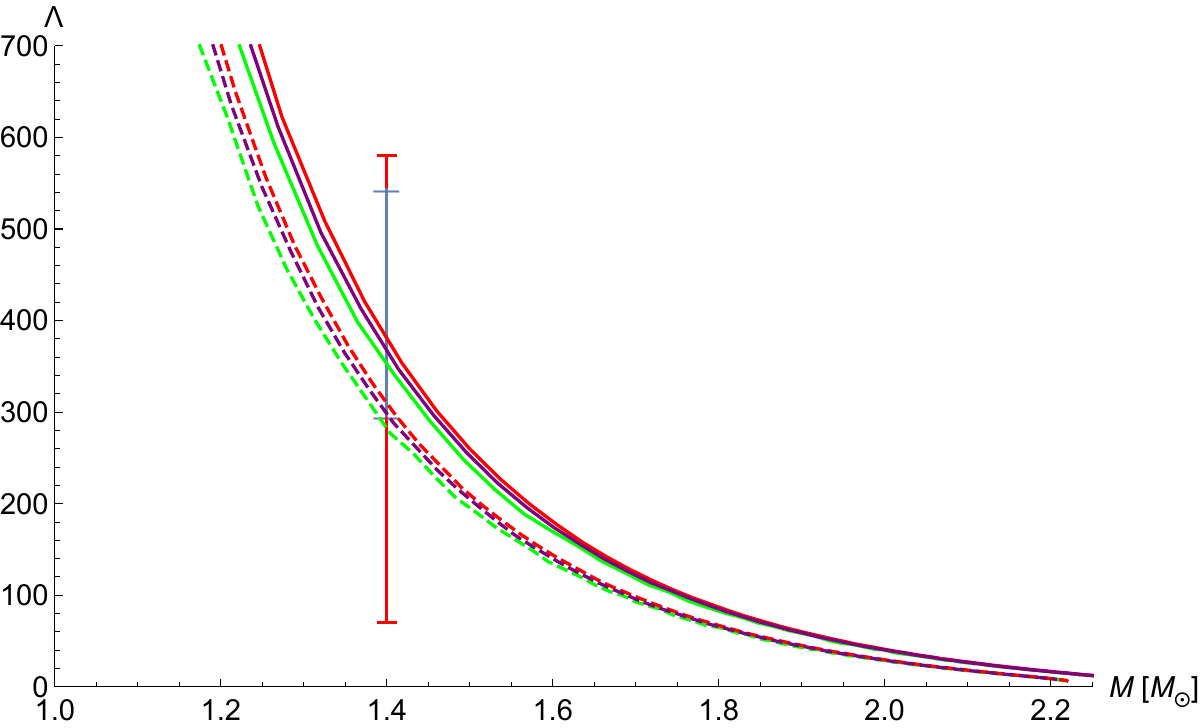}}
  \caption{Tidal deformability for the hybrid neutron stars of
    Fig.~\ref{fig:MR} for the two different fits. The color coding
    follows the same conventions as in Fig.~\ref{fig:MR}.
    The blue bound is from Ref.~\cite{Lim:2022fap}, whereas the red
    bound is from Ref.~\cite{LIGOScientific:2018cki}.
  }
  \label{fig:Tidal} 
\end{figure}

In Fig.~\ref{fig:Tidal} we plot the tidal deformability against the
mass of the star: the tidal deformability $\Lambda_{1.4}$ for a neutron star
of mass $1.4M_{\odot}$ ends up being in good agreement with currently established bounds
in the case of the fit corresponding to $S_N(d_0)=32.8\MeV$ for all three low-density EOSs, while the
other fit choice is either at tension with or slightly outside the bound from one of the references
adopted (despite both choices being in good agreement with the 
more loose bound from Ref.~\cite{LIGOScientific:2018cki}); another hint (together with the 
proton fraction at the junction density) that possibly
the favored fit is one closer to Eq.~\eqref{eq:fitLlambda}. 

The SLy4-WSS equation of state can be seen as one of intermediate
stiffness, between the soft and stiff CEFT ones: in particular, since
we used the extremal upper and lower bounds dictated by CEFT, and
since the junction density is determined to be lower than the value of
$1.1d_0$ employed in Ref.~\cite{Hebeler:2013nza} as the density at which
the equation of state changes to a polytropic piecewise expansion, we
conclude that these two curves also represent boundaries of a region
in the $M$-$R$ plane that encompass all possible hybridization of any
choice of low-density equation of state, hybridized with our
particular WSS one. Thus, we conclude that the consistency of the $M$-$R$
curves with current observations is a robust feature of our
holographic hybrid equation of state: of course the situation can
change once we adopt refinements to the hybridization process (such as
ones mentioned in the previous section), so we restrict ourselves to
only remark that, in the presence of these quite crude approximations,
the WSS model can provide an EOS for densities above saturation that
succeeds in reproducing neutron stars phenomenology.

\section{Conclusion}\label{sec:conclusion}

In this paper, we have computed neutron star mass-radius curves for
stable static stars in the Witten-Sakai-Sugimoto model.
Due to the approximation of using the homogeneous Ansatz, we
acknowledge that the results are not expected to be trustable at very
low densities below saturation density.
We have thus constructed a hybrid equation of state by patching the
SLy4 EOS and two other EOSs, namely the stiff and soft limit coming from CEFT,
together with that of the holographic WSS model, taking into
account the symmetry energy, charge neutrality and
$\beta$-equilibrium.
The calibration of the model in this paper, has been chosen by fixing
the saturation density to the physical one; this fixed the mass scale
$M_{\rm KK}$ of the model and furthermore fixed the symmetry energy
to the phenomenologically accepted value at saturation density, by
adjusting the 't Hooft coupling.
We have calculated everything for two fits, corresponding to the upper
(solid lines) and lower (dashed lines) value of the error bar on the
symmetry energy.
The results we find are unexpectedly good for a top-down holographic
QCD model.
In particular, we find a large maximal mass between $2.26M_\odot$ and
$2.35M_{\odot}$, which is hard to achieve for many nuclear physics and
phenomenologically driven models.
In addition, we pass the current mass-radius constraints as well as
the constraints on the tidal deformability (the latter only for one of
the fits).

The are many small improvements that could be made, in order to
further refine the results and hence to achieve precision neutron star
phenomenology from a top-down holographic QCD (WSS) based model.
In particular, we have not included the strange quark (s) in the
model, but since we also have not included quark masses, those should
be incorporated too, in order to achieve a physical model.
Furthermore, we have considered the hybrid model of patching together
the SLy4 with our holographic WSS results, but the low-density
equation of state could in principle be derived from the WSS model by
taking into account pasta phases, which in turn would require many
more computations such as wall tensions etc.
Another refinement of the model would be to take into account
higher-order corrections due to the isospin asymmetry, which would
correspond to $(\bchi\cdot\bchi)^2$ terms.
In principle, the WSS model itself also predicts more accurate
higher-order corrections, one example is to consider the full
Dirac-Born-Infeld action instead of the leading Yang-Mills term and
there are also corrections in $1/N_c$ and $1/\lambda$, although
computing those may be unpractical.

On top of these minor refinements, there are also more substantial
improvements we can make, that involve changing considerably our construction.
These are aimed at removing qualitative flaws, which we briefly review here, 
while at the same time providing proposals to remove or at least reduce 
each of the problems:
\begin{itemize}
\item The WSS model only has two free parameters in this configuration:
  this means that if we fit saturation density and symmetry energy, the chemical
  potential at the baryon onset is determined. It happens that these choices and the 
  configuration of the homogeneous Ansatz largely overestimate the value of the chemical potential at said onset, introducing the horizontal 
  segment in the EOSs we constructed. We can try to ameliorate this issue by refining the
  description of nuclear matter as the density decreases towards saturation, for example
  putting solitonic baryons on a lattice. At the same time, we should change the values of the 
  fit parameters to accommodate for the changes introduced. It is not clear if these two 
  effects will lower the value of the chemical potential, but it seems natural to consider
  a configuration with resolved individual baryons at such densities: it is after all exactly 
  at the boundary of the region within which we can deem the homogeneous approximation to be reliable.
\item The speed of sound exhibits a sharp discontinuity at the junction density: this 
  is not a realistic feature, especially since the transition density is very close to saturation. Around saturation density, the speed of sound is expected to be considerably lower 
  than the values we obtain: following the same reasoning as for the previous point, it is 
  reasonable to attribute this unphysical behavior to the inadequacy of the homogeneous Ansatz at
  saturation density. The introduction of an intermediate lattice phase could help with this issue, 
  even though it is highly unlikely that we can reach a continuous speed of sound with 
  the hybrid approach.
\item The speed of sound keeps growing monotonically until the TOV densities and
  beyond for a wide range of densities. It is expected that at the TOV scale, matter in neutron 
  stars exhibit deconfinement, hence being in a quark phase. This is a possibility we did not 
  explore in this work, and is certainly an interesting direction for future work: in 
  particular, the TOV masses we find are in the allowed region, but rather large for a static object, 
  and introducing the effects of a high-density quark phase could potentially lower the TOV mass.
  The introduction of such a phase requires a major change in the setup of the model: in particular,
  since the introduction of the backreaction from the baryons on the geometry seems an extremely 
  difficult problem to solve, the quark phase can be modeled instead in the non-antipodal setup
  of the flavor branes, working in the deconfined geometry in the decompactified limit. It is not
  clear to us at the moment if such a substantial change can
  be performed while preserving the nice agreement with observations that
  we have obtained in the present work.
\end{itemize}

We keep these three points as a guideline for future works.

\subsection*{Acknowledgments}

The work of L.~B.~is supported by the National Natural Science
Foundation of China (Grant No.~12150410316). 
S.~B.~G.~thanks the Outstanding Talent Program of Henan University and
the Ministry of Education of Henan Province for partial support.
The work of S.~B.~G.~is supported by the National Natural Science
Foundation of China (Grants No.~11675223 and No.~12071111) and by the
Ministry of Science and Technology of China (Grant No.~G2022026021L).

\bibliographystyle{JHEP}
\bibliography{bib}

\end{document}